\documentclass[aps,twocolumn,showpacs,superscriptaddress]{revtex4}
\usepackage{amsmath}
\usepackage{amssymb}
\usepackage{epsfig}
\usepackage{color}
\usepackage{graphicx,amsmath}
\usepackage[colorlinks]{hyperref}
\UseRawInputEncoding\input
\hypersetup{colorlinks,citecolor=red,linkcolor=blue,urlcolor=blue}
\usepackage{graphicx,amsmath}

\usepackage{bm}

\begin{document}

\title{Transport properties of strongly correlated Fermi systems}
\author{Vasily R. Shaginyan}\email{vrshag@thd.pnpi.spb.ru}
\affiliation{Petersburg Nuclear Physics Institute, NRC Kurchatov
Institute, Gatchina, 188300, Russia}\affiliation{Clark Atlanta
University, Atlanta, GA 30314, USA}\author{Alfred Z. Msezane}
\affiliation{Clark Atlanta University, Atlanta, GA 30314, USA}
\author{Mikhail V. Zverev} \affiliation{NRC Kurchatov
Institute, Moscow, 123182, Russia} \affiliation{Moscow Institute of
Physics and Technology, Dolgoprudny, Moscow District 141700,
Russia}

\begin{abstract}
Physicists are actively debating the nature of the quantum critical
phase transition that determines the low-temperature properties of
metals with heavy fermions. Important experimental observations of
their transport properties incisively probe the nature of the
quantum critical phase transition. In our short review, we consider
the transport properties of strongly correlated Fermi systems like
heavy fermion metals and high-$T_c$ superconductors. Their
transport properties are defined by strong inter-particle
interaction forming flat bands in these compounds. These properties
do not coincide with those of conventional metals. Indeed, in
contrast to the behavior of the transport properties of
conventional metals, the strongly correlated compounds exhibit the
linear in temperature resistivity, $\rho(T)\propto T$. We analyze
the magnetoresistance and show that it under the application of
magnetic field becomes negative. It is shown that near a quantum
phase transition, when the density of electronic states diverges,
semiclassical physics remains applicable to describe the
resistivity $\rho$ of strongly correlated metals due to the
presence of a transverse zero-sound collective mode, representing
the phonon mode in solids. We demonstrate that when $T$ exceeds the
extremely low Debye temperature $T_D$, the resistivity $\rho(T)$
changes linearly with $T$, since the mechanism of formation of the
$T$-dependence $\rho(T)$ is similar electron-phonon mechanism,
which predominates at high temperatures in ordinary metals. Thus,
in the region of $T$-linear resistance, electron-phonon scattering
leads to a lifetime of $\tau$ quasiparticles practically
independent of the material, which is expressed as the ratio of the
Planck constant $\hbar$ to the Boltzmann constant constant $k_B$,
$T\tau\sim \hbar/k_B$. We explain that due to the non-Fermi-liquid
behavior the real part of the frequency-dependent optical
conductivity $\sigma^R_{opt}(\omega)$ exhibits a scaling behavior,
and demonstrates the unusual power law behavior
$\sigma^R_{opt}(\omega)\propto\omega^{-1}$, rather than the
well-known one shown by conventional metals,
$\sigma^R_{opt}(\omega)\propto\omega^{-2}$. All our theoretical
considerations are illustrated and compared with the corresponding
experimental facts. Our results are in a good agreement with
experimental observations.
\end{abstract}


\maketitle

\section{Introduction}

An explanation of the rich and striking behavior of strongly
correlated electron liquid in heavy fermion (HF) metals,
high-temperature superconductors, quasicrystals, and etc is among
the main problems of condensed matter physics. Quantum phase
transitions (QPT) define the non-Fermi liquid (NFL) low-temperature
properties of strongly correlated Fermi systems. Their behavior in
the NFL state is so radical that the traditional quasiparticle
Landau paradigm cannot describe it. The underlying nature of QPT
continues to challenge theoretical understanding. Attempts have
been made to use concepts such as the Kondo lattice and quantum and
thermal fluctuations in QPT \cite{stew,loh,si,sach,col}. Alas, when
these approaches are designed to describe one property considered
central, they cannot explain others, even the simplest ones, such
as the Kadowaki-Woods relation \cite{kadw,shagrep}. This
relationship, which naturally arises with the leading role of
quasiparticles of the effective mass $M^*$, can hardly be explained
within the framework of a theory that assumes the absence of
quasiparticles, see e.g. \cite{shagrep,shag,atom,book_20}.
Arguments that quasiparticles in strongly correlated Fermi liquids
"get heavy and die" during QPT are usually based on the assumption
that the quasiparticle weight factor $Z$ vanishes at the point of
the corresponding second-order phase transition \cite{col1,col2} .
However, this scenario does not correspond to experimental facts
\cite{khodz,clark10,shagrep}. Numerous experimental facts have been
discussed within the framework of such a concept, but how it can
quantitatively explain the physics of HF metals remains an open
question. The theory of fermion condensation was proposed and
developed, preserving quasiparticles. The fermion condensation (FC)
takes place at the topological fermion condensation quantum phase
transition (FCQPT), and leads to both flat bands, that have been
predicted in 1990 \cite{ks}, and to the unlimited growth of the
effective mass $M^*$; at the same time, the extensive research has
shown that this theory provides an adequate theoretical explanation
for the vast majority of experimental results with various HF
metals
\cite{ks,volovik,volov_gr,volovik1,volovik2,shagrep,shag,atom,book_20,bern,catal,Khod_2020}.
Unlike the Landau paradigm, which is based on the assumption that
$M^*$ is an approximately constant, in the FC theory, $M^*$
strongly depends on temperature $T$, applied magnetic field $B$ and
etc. It is important to note that the extended quasiparticle
paradigm has been introduced. The essential point is that, as
before, well-defined quasiparticles determine the thermodynamic and
transport properties of strongly correlated Fermi systems
\cite{shagrep}. Indeed, the width of quasiparticles tends to zero
at $T\to0$, and they are well-defined excitations up to $T\sim 100$
K \cite{khodel:1994,shagrep}. In fact, this observation is in
accordance with numerous experimental observations; for example,
the typical behavior of the heat capacity of HF metals defined by
quasiparticles is observed in wide range of temperatures, see e.g.
\cite{shagrep}. The dependence of the effective mass $M^*$ on $T$
and $B$ leads to both the observed NFL behavior and the restoration
of the Landau Fermi liquid behavior at low temperatures under the
application of magnetic fields \cite{shagrep,shag,atom,book_20}.
The most fruitful strategy for studying and uncovering the nature
of QPT is to focus on those properties that exhibit the most
dramatic deviations from the Landau Fermi liquid (LFL) behavior of
ordinary metals at low temperatures \cite{lanl,pines,trio}.

In our review we consider the transport properties that allow one
to disclose the nature of QPT governing the behavior HF metals. In
particular, measurements of the magnetoresistance clarifies the
dependence of the effective mass $M^*$ on applied magnetic field
$B$, since, in contrast to ordinary metals, the magnetoresistance
becomes negative under the application of $B$, see e.g.
\cite{pag1,steg,oesbs}. This point is considered in Sections
\ref{Intr} and \ref{MR}. Relationships between the NFL resistivity
$\rho(T)\propto T$ and so called the Planckian limit open new
possibilities to analyze the properties of QPT that governs the
transport properties of HF metals \cite{bruin,legr}, see Sections
\ref{Lin} and \ref{Plan}. Precise experimental measurements of the
optical conductivity of HF metals $\rm YbRh_2Si_2$ and $\rm
La_{2-x}Sr_xCuO_4$ have been carried out, see e.g.
\cite{homes,pasch} which probe the nature of their QPT. It was
discovered that at low temperatures the optical conductivity is
very different from the well-known optical conductivity of ordinary
metals \cite{homes,pasch} see Sections \ref{Lin} and \ref{Opt}.
Section \ref{Intr} we consider general properties of the effective
mass $M^*$ in magnetic fields. Section \ref{Concl} is devoted to
the main conclusion of our review.

\section{The behavior of the effective mass}\label{Intr}

We start with analyzing the scaling behavior of the effective mass
$M^*$ and the schematic $T-B$ phase diagram of HF metals based on
the homogeneous HF liquid, thereby avoiding complications
associated with the crystalline anisotropy of solids
\cite{shagrep,book_20}. Before the topological FCQPT, the
temperature and magnetic field dependences of the effective mass
$M^*(T,B)$ is governed by the Landau equation
\cite{trio,pines,lanl}
\begin{eqnarray}
\nonumber \frac{1}{M^*_{\sigma}(T,
B)}&=&\frac{1}{m}+\sum_{\sigma_1}\int\frac{{\bf p}_F{\bf
p}}{p_F^3}F_
{\sigma,\sigma_1}({\bf p_F},{\bf p}) \\
&\times&\frac{\partial n_{\sigma_1} ({\bf
p},T,B)}{\partial{p}}\frac{d{\bf p}}{(2\pi)^3}. \label{HC1}
\end{eqnarray}
where $F_{\sigma,\sigma_1}({\bf p_F},{\bf p})$ is the Landau
interaction, $p_F$ is the Fermi momentum, and $\sigma$ is the spin
label.  {We note that Eq. \eqref{HC1} is an exact one, as it can be
shown within the framework of the Density Function Theory, see e.g.
\cite{shagrep,book_20,DFT}. } To simplify matters, we ignore the
spin dependence of the effective mass, noting that $M^*(T,B)$ is
nearly independent of spin in weak fields. The quasiparticle
distribution function $n({\bf p},T)$ is given by
\begin{equation} n_{\sigma}({\bf p},T)=\left\{ 1+\exp
\left[\frac{(\varepsilon({\bf p},T)-\mu_{\sigma})}T\right]\right\}
^{-1},\label{HC2}
\end{equation}
where $\varepsilon({\bf p},T)$ is the single-particle spectrum. In
the case being considered, the spectrum depends on spin only
weakly. However, the chemical potential $\mu_{\sigma}$ depends
non-trivially on spin due to the Zeeman splitting,
$\mu_{\pm}=\mu\pm B\mu_B$, where $\pm$ corresponds to states with
the spin ``up'' or ``down.'' Numerical and analytical solutions of
this equation show that the dependence of the effective mass
$M^*(T,B)$ on the temperature $T$ and magnetic field $B$ leads to
the appearance of three different regimes with increasing
temperature. The Fermi---Dirac distribution function can presented
as follows
\begin{equation}
\varepsilon({\bf p})-\mu=T\ln {1-n({\bf p})\over n({\bf p})},
\label{tem}
\end{equation}
where $\mu$ is the chemical potential and $n({\bf p})$ is the
quasiparticle occupation number. In the theory of fermion
condensation, if the system is located near the topological FCQPT
on its ordered side, the quasiparticle occupation number loses its
temperature dependence at sufficiently low $T$
\cite{ks,khodel:1994,shagrep}. In the interval $p_i\leq p\leq p_f$
the quasiparticle distribution function $n({\bf p})<1$, therefore
the logarithm in Eq. \eqref{tem} is finite and at $T=0$ the product
on the right hand side of Eq. \eqref{tem} is zero. As a result, on
the ordered side of the topological FCQPT the spectrum contains a
flat band \cite{ks,khodel:1994,shagrep}
\begin{equation}
\varepsilon({\bf p})-\mu=0\,\, {\rm if}\,\, p_i\leq p\leq p_f.
\label{tem1}
\end{equation}
 {The existence of the solution of
Eq. \eqref{tem1} means that the single-particle spectrum
$\varepsilon({\bf p},T)$  has a flat band. Since $\varepsilon({\bf
p})=\mu$ in the range $p_i\leq p\leq p_f$, the Fermi surface
spreads into a Fermi band, that is in the case of the three
dimensional Fermi sphere the two dimensional Fermi surface
transforms into three dimensional structure. Obviously, this
transformation gives rise to a change in the topological structure
of the single particle Green function, which makes us refer to the
systems with FC as a new class of Fermi liquids being different
from both the Landau Fermi liquid \cite{lanl} and the marginal
Fermi liquid, see e.g. \cite{marg}, and  with its specific
topological charge \cite{volovik,volovik2}. We note that Eqs.
\eqref{tem} and \eqref{tem1} are exact, see e.g.
\cite{shagrep,book_20,DFT}}

It is seen from Eq. \eqref{tem} that at any finite temperature the
flat band given by Eq. \eqref{tem1} vanishes and the effective mass
becomes finite \cite{ckhz,shagrep}. On the disordered side, at
finite $B$ and sufficiently low temperatures $T$ we have the LFL
state with $M^*(T)\simeq M^*+aT^2$, where $a$ is a positive
constant. Thus, the effective mass grows as a function of $T$,
reaching a maximum $M^*_M$ at a certain temperature $T_M$ and then
decreasing, see, for example \cite{shagrep,ckhz,oes}
\begin{equation}\label{r2}
M^*(T) \propto T^{-2/3}.
\end{equation}
\begin{figure} [! ht]
\begin{center}
\vspace*{-0.2cm}
\includegraphics [width=0.47\textwidth]{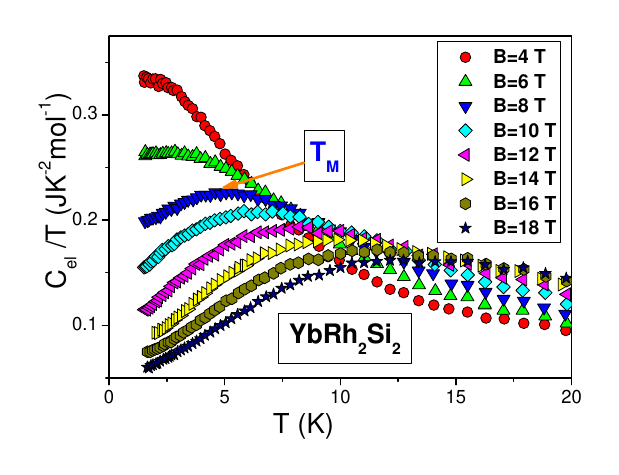}
\end{center}
\vspace*{-0.3cm} \caption{Electronic heat capacity $\rm
YbRh_2Si_2$, $C_{el}/T$, as a function of temperature $T$ and as a
function of magnetic field $B$ \cite{oes}, shown in the legend. As
an example, the maximum value $M_M$ taking place at $T_M$ and $B=8$
T is shown by the arrow}\label{fig1aa}
\end{figure}
The application of magnetic field restores the LFL behavior, and at
$T\leq T_M$ the effective mass depends on $B$ as \cite{shagrep}
\begin{equation}\label{B32}
M^*(B)\propto (B-B_{c0})^{-2/3},
\end{equation}
as it is seen from Fig. \ref{fig1aa}. Note that in some cases the
critical magnetic field $B_{c0}$ that tunes HF metal to its FCQPT
can be zero, $B_{c0}=0$. For instance, the HF metal
$\rm{CeRu_2Si_2}$ is characterized by $B_{c0}=0$ and shows no signs
of magnetic ordering, superconductivity, or the LFL behavior down
to the lowest temperatures \cite{takah}. Moreover, the closer the
control parameter $B$ is to its critical value $B_{c0}=0$, the
higher the growth rate. In this case, the peak value of $M^*_M$
also increases, but the temperature $T_M$, at which $M^*$ reaches
its maximum value, decreases, and $M^*_M(T_M,B\to
B_{c0})\to\infty$. At $T>T_M$ the LFL behavior disappears. When the
system is near FCQPT, the approximate interpolation solution to the
equation ~\eqref{HC1} has the form \cite{shagrep}
\begin{equation}
\frac{M^*}{M^*_M}={M^*_N(T_N)}\approx
c_0\frac{1+c_1T_N^2}{1+c_2T_N^{8/3}}. \label{UN2}
\end{equation}
Here, $T_N=T/T_M$ is the normalized temperature, with
$c_0=(1+c_2)/(1+c_1)$ in terms of fitting parameters $c_1$ and
$c_2$.  Since the magnetic field enters Eq.~\eqref{HC2} in the form
$\mu_BB/T$, we conclude that
\begin{equation}\label{YTB}
T/T_M=T_N\propto \frac{T}{\mu_BB},
\end{equation}
where $\mu_B$ is the Bohr magneton. It follows from Eq.~\eqref{YTB}
that
\begin{equation}
\label{TMB} T_M\simeq a_1\mu_BB.
\end{equation}
As a result, we conclude from Eqs. \eqref{YTB} and \eqref{TMB} that
$M_N(y)$ exhibits the scaling behavior as a function of both the
variables $y=T/B$ and $y=B/T$. Equation~\eqref{UN2} reveals the
scaling behavior of the normalized effective mass $M^*_N(T_N=y)$:
Values of the effective mass $M^*(T,B)$ at different magnetic
fields $B$ merge into a single mass value $M^*_N$ in terms of the
normalized variable $T_N\propto T/B\propto B/T$
\cite{shagrep,book_20}. Figure \ref{fig1} demonstrates the scaling
behavior of the normalized effective mass $M^*_N$ versus the
normalized temperature $T_N$. The LFL phase prevails at $T\ll T_M$,
followed by the $T^{-2/3}$ regime at $T \gtrsim T_M$. The latter
phase is designated as NFL due to the strong dependence of the
effective mass on temperature. The temperature region $T\simeq T_M$
covers the transition between the LFL regime with almost constant
effective mass and the LPL behavior described by the
equation~\eqref{r2}. Thus, $T\sim T_M$ defines a transition region
characterized by the intersection of the LFL and NFL regimes. The
inflection point $T_{\rm inf}$ of $M_N^*$ versus $T_N$ is depicted
by arrow in Fig.~\ref{fig1}.

\begin{figure}[!ht]
\begin{center}
\includegraphics [width=0.48\textwidth]{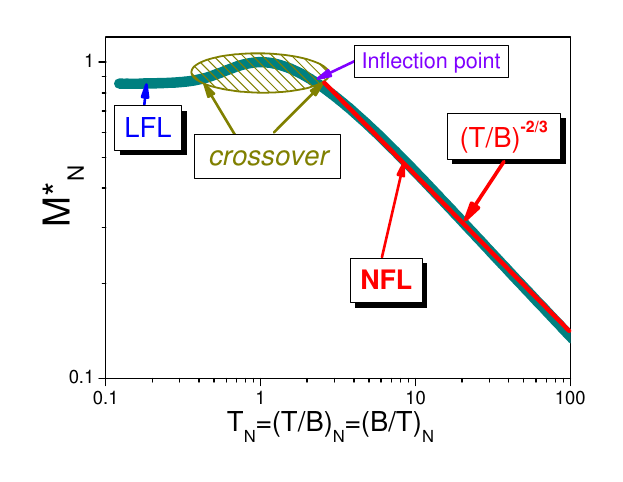}
\vspace*{-1.5cm}
\end{center}
\caption{The schematic plot of the normalized effective mass versus
the normalized temperature. The transition regime, where $M^*_N$
reaches its maximum value at $T_N=T/T_M=(T/B)_N=(B/T)_N=1$, is
shown as the hatched area. Arrows indicate the LFL region,
transition one, inflection point $T_{\rm inf}$ and NFL behavior
with $M^*_N\propto (T/B)^{-2/3}$.}\label{fig1}
\end{figure}

\begin{figure}[!ht]
\begin{center}
\includegraphics [width=0.47\textwidth]{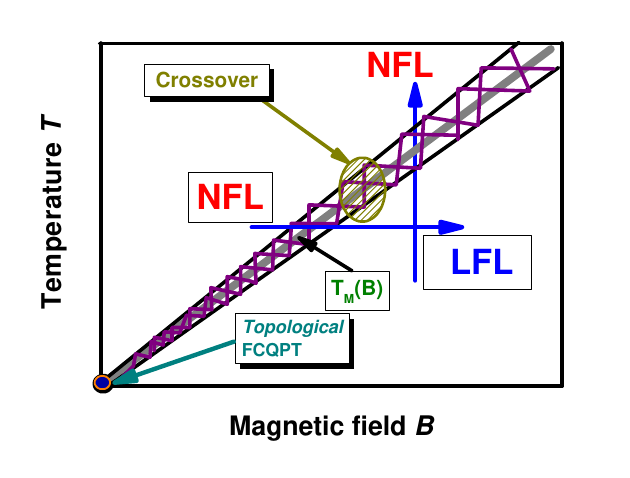}
\end{center}
\caption{Schematic $T-B$ phase diagram of a HF compound, with
magnetic field $B$ as control parameter.  The hatched area
corresponds to the crossover domain at $T_M(B)$. At fixed magnetic
field and elevated temperature (vertical arrow) there is a LFL-NFL
crossover. The horizontal arrow indicates a NFL-LFL transition at
fixed temperature and elevated magnetic field. The topological
FCQPT (shown by the arrow) occurs at $T=0$ and $B=0$.}\label{fig10}
\end{figure}
The transition (crossover) temperature $T_M(B)$ is not actually the
temperature of a phase transition.  Its specification is
necessarily ambiguous because it depends on the criteria used to
determine the point the crossover. Typically, the temperature
$T^*(B)$ is obtained from the field dependence of charge transfer,
for example, from the resistivity $\rho(T)$, determined by the
expression
\begin{equation}
\rho(T)=\rho_0+AT^{\beta},\label{res}
\end{equation}
where $\rho_0$ is the residual resistivity and $A$ is a
$T$-independent coefficient. The term $\rho_0$ is ordinarily
attributed to impurity scattering. The LFL state is characterized
by the $T^{\beta}$ dependence of the resistivity with index
$\beta=2$. The schematic phase diagram of a HF metal is depicted in
Fig.~\ref{fig10}, with the magnetic field $B$ serving as the
control parameter. The crossover (through the transition regime
shown as the hatched area in Fig.~\ref{fig10}) takes place at
temperatures where the resistance starts to deviate from LFL
behavior, with the exponent $\beta$ shifting from 2 into the range
$1<\beta<2$.  At $B=0$,  HF metal acquires a flat band
corresponding to a strongly degenerate state.

The NFL mode reigns at elevated temperatures and a fixed magnetic
field. As $B$ increases, the system moves from the NFL region to
the LFL domain. As shown in Fig.~\ref{fig10}, the system moves from
NFL mode to LFL mode by the horizontal arrow, and from LFL mode to
NFL mode by the vertical arrow. The magnetic field tuned QCP is
indicated by an arrow and is located at the beginning of the phase
diagram, since the application of a magnetic field destroys the
flat band and transfers the system to the LFL state
\cite{shagrep,shag,atom,book_20}. The shaded area, denoting the
transition region, separates the NFL state from the weakly
polarized LFL state and contains the dashed line displaying
$T_M(B)$.  Referring to equation~\eqref{TMB}, this line is defined
by $T=a_1\mu_BB$.  {It is worth noting that that the transition
from the NFL behavior the LFL taking place under application of
magnetic field, as it is seen from Figs. \ref{fig1} and
\ref{fig10}, is the special property of the topological FCQPT and
described by Eq. \eqref{UN2} \cite{shagrep,shag_pla}. This
important property is in a good agreement with experimental facts
and allows one to use it as the versatile tool to explore the
physics of HF compounds, including the violation of both the
particle-hole symmetry and the time invariance symmetry; this
violation is directly related to the concept of flat bands
\cite{shagrep,mdpi23}, see Sections \ref{Plan} and \ref{Opt}. In
contrast, these properties are not considered in a number of
theories, including the theory of marginal Fermi liquid, see e.g.
\cite{marg}.}

\section{Longitudinal magnetoresistance} \label{MR}

\begin{figure} [! ht]
\begin{center}
\vspace*{-0.5cm}
\includegraphics [width=0.47\textwidth]{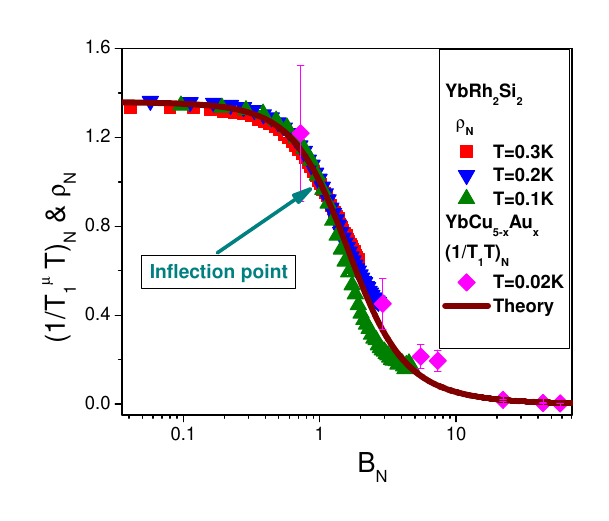}
\end{center}
\vspace*{-0.8cm} \caption{Magnetic field dependence of both the
normalized magnetoresistance $\rho_N$ and the muon spin-lattice
relaxation rate $(1/T_1^\mu T)_N$ versus normalized magnetic field
$B_N$. $\rho_N$ was extracted from LMR of $\rm YbRh_2Si_2$ at
different temperatures \cite{steg,oesbs} listed in the legend.
Magnetic field dependence of normalized muon spin$-$lattice
relaxation rate $1/T_1^\mu T$ in ${\rm {YbCu_{4.4}Au_{0.6}}}$ is
shown by diamonds and extracted from \cite{osn}. The data are
normalized in the inflection point and shown by the arrow. The
solid line represents our calculations, see Eqs. \eqref{B32},
\eqref{rn} and \eqref{chi5ab}.}\label{fig4H}
\end{figure}

Consider a longitudinal magnetoresistance (LMR)
\begin{equation}\label{KWR}
\rho(B,T)=\rho_0+A(B)T^2,
\end{equation}
as a function of $B$ at fixed $T$. In that case, the classical
contribution to LMR formed by orbital motion of carriers induced by
the Lorentz force is small. In the LFL state, the Kadowaki-Woods
relation is given by \cite{kadw,shagrep}
\begin{equation}\label{KW}
K=A/\gamma_0^2\propto A/\chi^2=const,
\end{equation}
allows us to employ $M^*$ to construct the coefficient $A$, since
$\gamma_0\propto\chi\propto M^*$. Here $\gamma_0$ is the Sommerfeld
coefficient and $\chi$ is the magnetic susceptibility. Omitting the
classical contribution to LMR, we obtain that
$\rho(B,T)-\rho_0\propto(M^*)^2$ \cite{shag_pla}. The magnetic
field dependence of the muon spin-lattice relaxation rate
$1/T_1^\mu$ is given by \cite{shag_pla,osn}
\begin{equation}\label{chi5}
\frac{1}{T_1^\mu T}=\eta\left[M^*(T,B)\right]^2,
\end{equation}
where $\eta$ is a const. We note here that experimentally observed
relation
\begin{equation}\label{chi5a}
\frac{1}{T_1^\mu T}\propto \chi^2
\end{equation}
follows explicitly from Eqs. \eqref{KW} and \eqref{chi5}
\cite{shag_pla}. Figure \ref{fig4H} shows the normalized values of
both the magnetoresistance of $\rm YbRh_2Si_2$  \cite{steg,oesbs}
\begin{equation}
\label{rn} \rho_N(B_N)=(M_N^*(B_N))^2
\end{equation}
and the muon spin-lattice relaxation rate of ${\rm
{YbCu_{5-x}Au_x}}$ (x=0.6) \cite{osn}
\begin{equation}
\label{chi5ab} \left(\frac{1}{T_1^\mu T}\right)_N=(M_N^*(B_N))^2
\end{equation}
versus normalized magnetic field $B_N=B/B_{inf}$ at different
temperatures, shown in the legend. It is seen from Eqs.
\eqref{B32}, \eqref{rn} and \eqref{chi5ab} that both LMR and the
the muon spin-lattice relaxation rate are diminishing functions of
magnetic field $B$. This result is the vivid feature of the fermion
condensation theory that allows one to evaluate the behavior the
effective mass under the application of magnetic fields, see e.g.
\cite{shagrep,mdpi23}.

The normalization procedure deserves a remark here. Namely, since
the magnetic field dependence (both theoretical and experimental)
of $1/T_1^\mu$ and LMR do not have "peculiar points" like extrema,
the normalization have been performed in the inflection point,
corresponding to maximum of corresponding derivative. It is seen
that such procedure immediately reveals the universal magnetic
field behavior of both the reciprocal relaxation time and LMR,
showing their proportionality to the effective mass square, see
Eqs. \eqref{rn} and \eqref{chi5ab}. This behavior obtained directly
from the experimental findings is a vivid evidence that the above
quantities behavior is predominantly governed by field $B$ and
temperature $T$ dependence of the effective mass $M^*(B,T)$ given
by Eq. \eqref{UN2}. We note that the entire field (and temperature)
dependence of both $1/T_1^\mu T$ and LMR is completely determined
by corresponding dependence of the effective mass $M^*_N$ shown in
Fig. \ref{fig1}. The fact that the effective mass becomes field
$B$, temperature $T$ and the other external parameters dependent is
the key consequence of the FC theory. Both the theoretical curve
and the experimental data have been normalized by their inflection
points, which also reveal the universal scaling behavior: The
curves at different temperatures merge into single one in terms of
scaled variable $B_N$. Figure \ref{fig4H} shows clearly that both
the normalized magnetoresistance $\rho_N$ and the muon spin-lattice
relaxation rate time $1/T_1^\mu T$ well obeys the scaling behavior
given by Eqs. \eqref{UN2}, \eqref{rn} and \eqref{chi5ab} and shown
in Fig. \ref{fig1}.

\section{Linear in temperature resistivity}\label{Lin}

To analyze the resistivity given by Eq. \eqref{res} at elevated
temperatures $T$ and under the application of magnetic field $B$,
we assume that the electron system of HF metal contains a flat
band. The flattening of the single-particle spectrum
$\varepsilon({\bf p})$ is directly related to the problem being
solved, since as a result of Umklapp processes, quasiparticles of
the flat band create a contribution to $\rho_0$ indistinguishable
from the contribution due to the scattering of impurities
\cite{kz}. Furthermore, it is crucial that the flat band somehow
becomes destroyed at $T\to 0$ and under the application of magnetic
field the HF metal transits to the LFL state, see Fig. \ref{fig10}.
This destruction entails a strong suppression of the flat-band
contribution to $\rho_0$ \cite{kz}. Before proceeding to the
analysis of this destruction, we pay attention to vivid
consequences of the flattening of $\varepsilon({\bf p})$ in
strongly correlated Fermi systems. The theoretical possibility of
this phenomenon and its consequences, also called swelling of the
Fermi surface or the fermion condensation, was discovered a few
decades ago \cite{ks,volovik,noz}, for recent reviews, see e.g.
\cite{shagrep,shag,atom,book_20}. At $T=0$ the ground state of the
flat-band system is degenerate, and so the occupation number
$n_0({\bf p})$ of single-particle states belonging to the flat band
forming the fermion condensate are continuous functions of momentum
that interpolate between standard LFL values $\{0,1\}$ in the area
occupied by FC, see Eq. \eqref{tem1}. This leads to an entropy
excess
\begin{equation}
S_0=-\sum_{\,{\bf p}} n_0({\bf p})\ln n_0({\bf p})+(1-n_0({\bf
p}))\ln(1- n_0({\bf p})), \label{S*}
\end{equation}
which does not contribute to the specific heat $C(T)$.  It is seen
from Eq. \eqref{S*} that in contrast to the corresponding LFL
entropy vanishing linearly as $T\to 0$ the entropy of the system
with the fermion condensate $S(T\to0)\to S_0$. In the theory of
fermion condensation, the aforementioned ground-state degeneracy is
lifted at any finite temperature, where FC acquires a small
dispersion proportional to $T$, see Eq. \eqref{tem}. However, the
removal of degeneracy with increasing temperature does not change
the occupation number $n_0({\bf p})$, which means that the excess
entropy $S_0$ will persist down to zero temperature. To avoid a
subsequent violation of Nernst's theorem, it is necessary to
completely eliminate FC at $T\to 0$. In the most natural scenario,
this occurs through a SC phase transition, in which FC is destroyed
with the appearance of a $\Delta$ pairing gap in the
single-particle spectrum \cite{ks,khodel:1994,shagrep}. We assume
that this scenario is realized in CeCoIn$_5$ at sufficiently weak
magnetic fields, ensuring the elimination of the flat portion in
the spectrum $\varepsilon({\bf p})$ and the removal of excess
entropy $S_0$  \cite{kz}. In stronger external magnetic fields $B$
sufficient to terminate superconductivity in $\rm CeCoIn_5$, this
path becomes ineffective, giving way to an alternative scenario
involving a transition from the FC state to the LFL state with a
multiply connected Fermi surface \cite{shagrep}. In the phase
diagram depicted in Fig. \ref{fig10}, such a crossover is indicated
by the hatched area between the domains of NFL and LFL behavior and
also by the line $T_{\rm M}(B)$. In case of the HF metal CeCoIn$_5$
the end point of the curve $T_{\rm M}(B)$ nominally separating NFL
and LFL phases is the magnetic field inducing the topological FCQPT
hidden in the SC state \cite{kz,ronn,oeschler}. This is the most
characteristic feature of the phase diagram of the behavior of
resistivity $\rho(T,B)$. Since the entropies of the two phases are
different, near the topological FCQPT the SC transition should
become of the first order \cite{shagrep}, which is consistent with
the experimental fact \cite{bianchi}. Moreover, under the
application of sufficiently high magnetic field $B$, the LFL
behavior remains in effect even ta $T\to 0$. Thus, the imposition
of magnetic field $B$ drives the system in question from its SC
phase to the LFL phase. As a result, the  FC state or equivalently
the flat portion of the spectrum $\varepsilon({\bf p})$, is
destroyed. Thus, application of a high magnetic field to CeCoIn$_5$
is to cause a step-like drop in its residual resistivity $\rho_0$,
as it is seen experimentally \cite{pag1}. In addition, it should be
expected that the higher the quality of the CeCoIn$_5$ single
crystal, the stronger the suppression of $\rho_0$. Now we consider
the low-temperature transport properties of the normal state of
CeCoIn$_5$. We use a two-band model, one of which is assumed to be
flat with dispersion given by  equation ~\eqref{tem}, and the
second band is assumed to have a single-particle LFL spectrum with
finite $T$-independent dispersion \cite{kz}.

We begin our consideration with the case when a HF metal in its
normal state, where the resistivity is a linear function of $T$.
This behavior is inherent in electronic systems with flat bands.
Now we deifine the conductivity $\sigma(T)$ in terms of the
imaginary part of the polarization operator $\Pi({\bf j})$
\cite{trio},
\begin{eqnarray} \nonumber
\sigma&=&\lim \omega^{-1}{\rm Im} \Pi({\bf j},\omega\to 0)\propto
{1\over T}\int\int {d\upsilon d\epsilon \over
\cosh^2(\epsilon/2T)}\\
&\times&|{\cal T}({\bf j},\omega=0)|^2{\rm Im} G_R({\bf
p},\epsilon){\rm Im} G_R({\bf p},\epsilon),\label{con1}
\end{eqnarray}
where $d\upsilon$ is an element of momentum space, ${\cal T}({\bf
j},\omega)$ is the vertex part, ${\bf j}$ is the electric current,
and $G_R( {\bf p},\epsilon)$ is the retarded quasiparticle Green
function. The imaginary part reads
\begin{equation}
{\rm Im} G_R({\bf p},\epsilon)= -{\gamma\over
(\varepsilon-\epsilon({\bf p}))^2+\gamma^2} \label{gr}
\end{equation}
in terms of the spectrum $\varepsilon({\bf p})$ and the damping
$\gamma$ related to the band with a finite value $v_F$ of the Fermi
velocity. Applying gauge invariance, we get ${\cal T}({\bf
j},\omega=0)=e\partial\varepsilon({\bf p})/\partial {\bf p}$ .
Substituting this equation into Eq.~\eqref{con1} and doing some
algebra, we arrive at the standard result\begin{equation}
\sigma(T)= e^2n{v_F\over \gamma(T)}, \label{con}
\end{equation}
where $n$ is the number density of electrons.
\begin{figure}[!ht]
\begin{center}
\includegraphics [width=0.47\textwidth]{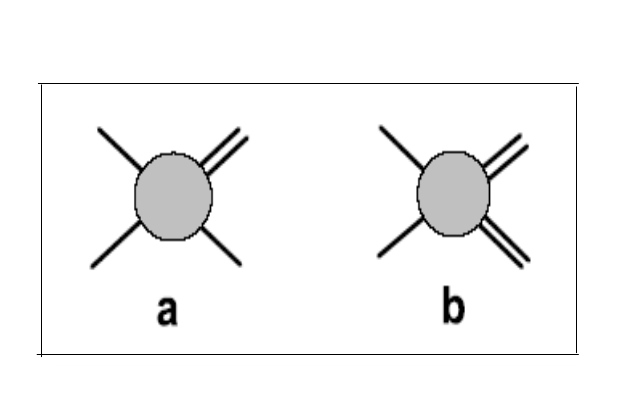}
\end{center}
\caption{ Scattering diagrams contributing to the imaginary part
mass operator $\Sigma(\epsilon)$ related to the band with a finite
value of the Fermi velocity $p_F$. The single line corresponds to a
quasiparticle of this band, the double line to the FC
quasiparticle.} \label{fig6}
\end{figure}
In ordinary pure metals obeying the LFL theory, the damping
$\gamma(T)$ is proportional to $T^2$, which leads to the Eq.
~\eqref{res} with $\beta=2$. The NFL behavior of $\sigma(T)$ is due
to the NFL dependence of $\gamma(T)$ on temperature associated with
the presence of FC \cite{kz}. In a standard situation, when the
volume $\eta$ occupied by FC is quite small, the overwhelming
contribution to the transport is made by inelastic scattering,
schematically presented in Figs. \ref{fig6}a and  \ref{fig6}b,
where FC quasiparticles (highlighted by a double line) turn into
normal quasiparticles; or vice versa, normal quasiparticles rotate
into FC quasiparticles. The contributions of these processes to
damping $\gamma$ are estimated based on the simplified Equation:
\cite{trio}
\begin{equation}
\nonumber \gamma({\bf
p},\varepsilon)\propto\int\int\int\limits_0^\epsilon
\int\limits_0^\omega |\Gamma({\bf p},{\bf p}_1,{\bf q}|^2{\rm Im}
G_R({\bf p}-{\bf q},\varepsilon-\omega)
\end{equation}
\begin{equation}\times{\rm Im} G_R( -{\bf p}_1,-\varepsilon)
{\rm Im} G_R({\bf q}-{\bf p}_1,\omega-\varepsilon) d{\bf p}_1 d{\bf
q} d\omega d\varepsilon, \label{gamma}
\end{equation}
where now the volume element in momentum space includes summation
over different bands. The straightforward calculations give:
\begin{equation}
\gamma(\epsilon)=\eta(\gamma_0+\gamma_1\epsilon),\quad {\rm
Re}\Sigma(\epsilon)=-\eta\gamma_1 \epsilon \ln {\epsilon_c\over
|\epsilon|}, \label{damp}
\end{equation}
where $\eta$ denotes the volume of momentum space occupied by the
flat band, and $\epsilon_c$ is the characteristic constant defining
the logarithmic term in $\Sigma$. Taking vertex corrections into
account \cite{trio} provides transparent changes to Eq.
\eqref{gamma} and cannot be held responsible for the effects
discussed here.
\begin{figure}
\begin{center}
\includegraphics [width=0.68\textwidth]{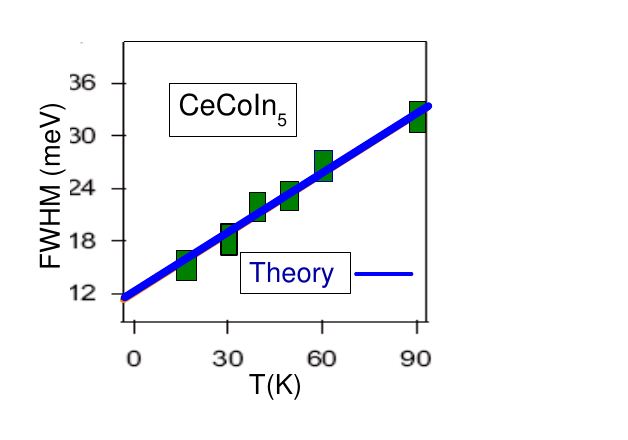}
\end{center}
\caption{The temperature dependence of the full width at half
maximum (FWHM) of the single-particle scattering rate of the main
Kondo resonance \cite{chen} is shown by solid squares. The line
represents the best fit: FWHM $= 11.8 + 2.69Tk_B$ meV, where $k_B$
is Boltzmann's constant \cite{chen}. The solid line is our
calculations \cite{kz}.}\label{fig9}
\end{figure}
Note that Eq. \eqref{dampp} gives the lifetime $\tau$ of
quasiparticles,
\begin{equation}
\hbar/\tau=\gamma(T)\simeq a_0+a_1T, \label{damT}
\end{equation}
where $\hbar$ is Planck's constant, $a_0$ and $a_1$ are parameters.
Combining Eqs. \eqref{damp} and \eqref{damT}, we obtain
\begin{equation}
\hbar/\tau(\epsilon,T)=\gamma(\epsilon,T)\simeq
a_0+a_1T+a_2\epsilon. \label{damtot}
\end{equation}
Where $a_0\propto\rho_0$, $a_1$ and $a_2$ are parameters. This
result is in good agreement with experimental facts
\cite{tomph,chen}, as it is seen from Fig. \ref{fig9}. Considering
Eq. \eqref{damT}, one immediately see that $\rho(T)=\rho_0+AT$,
i.e., the resistivity $\rho(T)$ of systems containing FC, is indeed
a linear function of $T$, which is consistent with experimental
data collected on $\rm CeCoIn_5$, see Fig. \ref{fig9}. Moreover,
the $\rho_0$ term appears even if the metal has an ideal lattice
and no impurities at all.

\section{$T$-linear resistivity and Planckian limit}\label{Plan}

\begin{figure} [! ht]
\begin{center}
\includegraphics [width=0.47\textwidth]{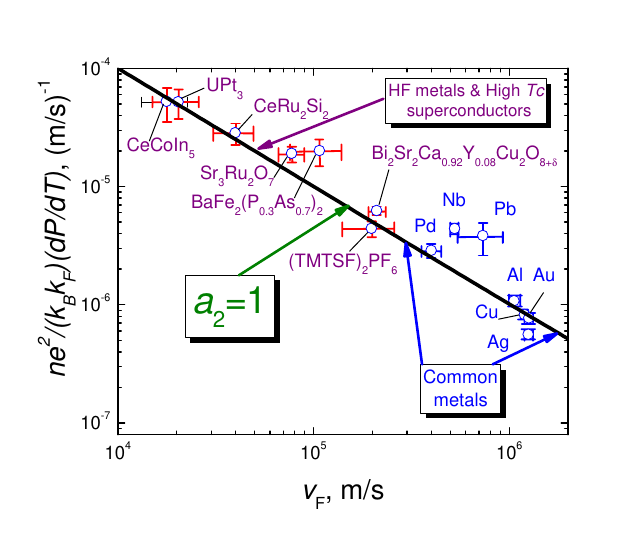}
\end {center}
\caption{ Scattering rates per kelvin vary greatly in correlated
metals such as HF, HTSC, organic and common metals \cite{bruin}.
All these metals have $\rho(T)\propto T$ and exhibit a change in
Fermi velocities $v_F$ by two orders of magnitude. The parameter
$a_2\simeq 1$ delivers the best fit, displayed by the solid line,
and corresponds to the scattering rate $\tau T=h/(2\pi k_B)$ with
$h=2\pi \hbar$, see Eqs. \eqref{vf} and \eqref{planc}. The region
occupied by the conventional metals is highlighted by two (blue)
arrows. The single (green) arrow shows the region of strongly
correlated metals, including organic ones. Note, that at low
temperatures $T\ll T_D$, the scattering rate per kelvin of a
conventional metal is orders of magnitude lower, and does not
correspond to the Planckian limit. The area occupied by ordinary
metals is highlighted by two (blue) arrows. The single (green)
arrow shows the region of strongly correlated metals, including
organic ones. We emphasize that at low temperatures $T\ll T_D$ the
scattering rate per kelvin of an ordinary metal is orders of
magnitude lower and is not in accordance with the Planck limit.}
\label{Sc1}
\end{figure}

The exotic experimentally observed properties of various classes of
HF compunds still remain largely unexplained due to the lack of a
universal underlying physical mechanism. These properties are
usually attributed to the so-called non-Fermi-liquid (NFL)
behavior. The latter behavior is widely observed in heavy fermion
(HF) metals, graphene, and high $T_c$ (HTSC) superconductors.
Experimental data collected at many of these systems show that at
$T=0$ some of them the excitation spectrum becomes dispersionless,
which leads to flat bands, see e.g.
\cite{ks,volovik,volovik1,prb:2013,cao}. The presence of flat band
indicates that the system is close to the topological
fermion-condensation quantum phase transition (FCQPT)
\cite{ks,volovik,volovik1,cao}, leading to a the formation of flat
band given by Eq. \eqref{tem1}. Vivid experimental data on the
linear temperature $T$ dependence of resistivity $\rho(T)\propto
T$, collected on HTSC, graphene, HF and ordinary metals, showed
that the charge carrier scattering rate 1$/\tau$ reaches the
so-called universal Planck limit $1/(T\tau)=k_B/\hbar$ ($k_B$ and
$\hbar=h/2\pi$ are the Boltzmann and Planck constants,
respectively) \cite{bruin,legr,cao}. Note that this is above the
Planck limit used to explain the universal scattering rate in
so-called Planck metals \cite{bruin,legr,cao} can happen by chance
because experimental manifestations in metals other than Planck can
be just as well explained by more traditional physical mechanisms,
such as those associated with phonon contributions\cite{prb:2013}.
For example, ordinary metals exhibit a universal linear scattering
rate at room and higher temperatures generated by well-known
phonons, which are classical lattice excitations \cite{bruin}. It
is shown that, within the framework of the theory of fermion
condensation, semiclassical physics is still applicable to describe
the universal scattering rate 1$/\tau$ experimentally observed in
strongly correlated metals in their quantum critical region. This
is due to the fact that the flat zones responsible for quantum
criticality generate a transverse zero-sound mode, reminiscent of
phonon mode in solids with Debye temperature $T_D$
\cite{prb:2013,lett2010}. At $T\geq T_D$, the mechanism of linear
temperature dependence of resistivity is the same in both ordinary
and strongly correlated metals, and is represented by
electron-phonon scattering. Consequently, it is the scattering of
electrons on phonons at $T\geq T_D$ that gives almost material
independence of the lifetime $\tau$. It is expressed as $1/(\tau
T)\sim k_B/\hbar$. Thus, the exciting experimental observations of
universal scattering rate related to linear-temperature resistivity
of a large number of both strongly correlated Fermi systems and
conventional metals can be explained
\cite{bruin,legr,cao,prb:2013,lett2010}. The observed scattering
rate is well explained by the appearance of flat bands formed by
the topological FQCPT, rather than by the so-called Planck limit at
which the supposed Planck scattering rate occurs. At low
temperatures, the observed  resistivity in their normal state both
HTSC and HF metals obeys linear law given by Eq. \eqref{res} with
$\beta=1$. On the other hand, at room temperature the $T$-linear
resistivity is exhibited by conventional metals such as $\rm Al$,
$\rm Ag$ or $\rm Cu$. In the case of a simple metal with a single
pocket on the Fermi surface, the resistivity has the form
$e^2n\rho=p_F/(\tau v_F)$, where $\tau$ is the lifetime, $e$ is the
electronic charge, $n$ is carrier concentration. The lifetime
$\tau$ (or inverse scattering rate) of quasiparticles can be
presented as
\begin{equation}\label{LT}
\frac{\hbar}{\tau}\simeq a_1+\frac{k_BT}{a_2},
\end{equation}
we obtain \cite{prb:2013}
\begin{equation}\label{vf}
a_2\frac{e^2n\hbar}{p_Fk_B}\frac{\partial\rho}{\partial
T}=\frac{1}{v_F},
\end{equation} where $a_1$ and $a_2$ are $T$-independent
parameters. There are two challenging points for a theory. The
first point is that experimental data confirm Eq. \eqref{vf} for
both strongly correlated metals (HF metals and HTSC) and ordinary
ones, provided that these demonstrate the linear $T$-dependence of
their resistivity \cite{bruin}, see Fig. \ref{Sc1}. The second
point is, that under the application of magnetic field, HF metals
exhibit the LFL behavior, see Fig. \ref{fig10}. For example, the HF
metal $\rm CeRu_2Si_2$ exhibits the LFL behavior under the
application of magnetic field as small as the magnetic field of the
Earth \cite{takah}.  {Obviously, both of these two facts cannot be
explained with the standard theories, see e.g.
\cite{sachp,volov19,marg}, since the ordinary metals have nothing
to do with the Planckian limit; moreover, such a small magnetic
field cannot destroy the limit, since the LFL behavior does not
related with the limit.}
The coefficient $a_2$ is always close to unity, $0.7\leq a_2\leq
2.7$, despite the huge difference in the absolute values of $\rho$,
$T$ and Fermi velocities $v_F$, which differ by two orders of
magnitude \cite{bruin}. As a result, it follows from Eq. \eqref{LT}
that the $T$-linear scattering rate is of universal form, $1/(\tau
T)\sim k_B/\hbar$. This takes place in different systems displaying
the $T$-linear dependence with parameter entering Eq. \eqref{vf},
$a_2\simeq 1$ \cite{bruin,prb:2013}. Indeed, such a dependence is
demonstrated by ordinary metals at temperatures above the Debye
one, $T\geq T_D$, with an electron-phonon mechanism, as well as
strongly correlated metals, that are supposed to be fundamentally
different from ordinary ones, in which the linear dependence at
their quantum criticality and temperatures of a few Kelvin are
assumed to be due to electronic excitations rather than phonons
\cite{bruin}. As can be seen from Fig. \ref{Sc1}, this scaling
relation spans two orders of magnitude in $v_F$, indicating the
stability of the observed empirical law \cite{bruin}. This behavior
is explained within the framework of the PC theory, since for both
ordinary and strongly correlated metals, the scattering rate is
determined by phonons \cite{prb:2013,lett2010}. In the case of
ordinary metals at $T>T_D$ it is well known that the main
contribution to the linear dependence of resistivity is made by
phonons. On the other hand, it is shown that semiclassical physics
describes the $T$-linear dependence of the electrical resistance of
strongly correlated metals at $T>T_D$, since the flat bands forming
quantum criticality generate a transverse zero-sound mode with the
Debye temperature $T_D$ located inside area of quantum criticality
\cite{prb:2013,lett2010}. Consequently, the $T$-linear dependence
is formed due to electron-phonon scattering both in ordinary metals
and in strongly correlated ones. Thus, it is electron-phonon
scattering that leads to almost material independence of the
lifetime $\tau$, which is expressed as
\begin{equation}\label{planc}
\tau T\sim \frac{\hbar}{k_B}.
\end{equation}
We emphasize that the Planck limit can arise by chance: it is
extremely unlikely that it will occur in ordinary metals, which
obviously cannot be recognized as Planck limit with quantum
criticality at high or low temperatures. The fact that we observe
the same universal scattering rate behavior in microscopically
different highly correlated compounds such as HTSC, HF and common
metals suggests that some general theory is needed to provide a
unified explanation for the above set of materials and their
behavior. We confidently conclude that FC theory is a responsible
approach to explaining the physics of strongly correlated Fermi
systems.

\section{The optical conductivity of heavy fermion metals}\label{Opt}

In this Section, we use the FC theory to explain the NFL behavior
of the optical conductivity based on experimental facts
\cite{homes,proch,mich,pasch}. We show that $\omega/T$-scaling
behavior of optical conductivity $\sigma_{opt}(\omega,T)$ is
exhibited by HF compounds, where $\omega$ and $T$ are frequency and
temperature, respectively. We show that because of the linear
temperature dependence of the electrical resistivity,
$\rho(T)\propto T$, and at $\omega/T\geq 1$, the real part
$\sigma^R_{opt}(\omega,T)$ of the optical conductivity
$\sigma_{opt}(\omega,T)$ demonstrates the unusual power law
behavior $\sigma^R_{opt}\propto\omega^{-1}$.

Modern condensed matter physics is vividly represented by the
experimental discovery of flat bands \cite{cao,bern,catal}, as they
were predicted many years ago
\cite{ks,volovik,volov_gr,khodel:1994,shagrep}. One can expect the
existence of a general physical mechanism generated by the presence
of flat bands and manifested in the universal scaling behavior of
HF compounds. Indeed, HF compounds do exhibit universal scaling
behavior and specific behavior caused by the presence of flat bands
\cite{shagrep,book_20,atom,Khod_2020}. Within the framework of the
fermion condensation theory such a mechanism is represented by the
topological FCQPT supporting quasiparticles, surviving the
unlimited growth of the effective mass $M^*$, forming the non-Fermi
liquid (NFL) behavior and generating flat bands
\cite{ks,volovik,khodel:1994,shagrep,book_20}. The main goal of the
quasiparticle interaction is to place the system at the topological
FCQPT. As a result, the universal scaling behavior of HF metals can
be explained, for it becomes independent of the interactions near
the formation of flat bands. Thus, the universal scaling becomes
independent of interaction strength and its other properties for
sufficiently large interactions \cite{shagrep,book_20}. However, it
is important to explore new properties of HF compounds that are not
directly determined by the effective mass $M^*$ and cannot be
explained within the framework of theories based on ordinary
quantum phase transitions, the Kondo-breakdown scenario, etc., see,
for example, \cite{homes,proch,mich,pasch}. For example, within the
framework of the FC theory, the linear temperature dependence of
electrical resistance $\rho(T)\propto T$ is explained, which is one
of the main features of the behavior of the NFL \cite{lett2010,kz},
and can lead to a special behavior of optical conductivity exhibit
HF metals. To analyze the optical conductivity
$\sigma_{opt}(\omega,T)$ we use the Drude model , see e.g.
\cite{pines},
\begin{equation}
\sigma_{opt}(\omega,T)=\sigma_0\frac{1}{1-i\omega\tau}.\label{DRUD}
\end{equation}
Here $\sigma_0$ reads
\begin{equation}
\sigma_0(T)=\frac{ne^2\tau(\omega,T)}{m}|_{\omega=0}, \label{DR1}
\end{equation}
where the lifetime $\tau$ is given by Eq. \eqref{damtot}
\cite{lett2010,kz,prb:2013}
\begin{equation}
\tau(\omega,T)=\frac{1}{a_0+a_1T+a_2\omega}.\label{DR2}
\end{equation}
Here $a_0\propto\rho_0$, where $\rho_0$ is the residual
resistivity; $m$, $a_1$ and $a_2$ are coefficients. The residual
resistivity includes two contributions,
$\rho_0=\rho_{imp}+\rho_{FC}$, where $\rho_{imp}$ comes from the
impurities that holds a HF metal and $\rho_{FC}$ is formed by the
FC state \cite{kz}, see Section \ref{Lin}. It is seen from Eqs.
\eqref{DR1} and \eqref{DR2} that
\begin{equation}
\rho(T)=\rho_0+\frac{m}{ne^2}\,a_1T. \label{DR3}
\end{equation}
The NFL behavior of the lifetime $\tau$ is given by Eq.
\eqref{DR2}, while in the LFL theory $\tau$ is given by
\cite{gurz,pines,lanl}
\begin{equation}
\tau(\omega,T)=\frac{1}{a_0+c_1T^2+c_2\omega^2}, \label{tu}
\end{equation}
where $c_1$ and $c_2$ are parameters and $a_0\propto\rho_{imp}$,
since FC is absent. As we will see, the NFL behavior of optical
conductivity is determined by the NFL dependence of $\tau$ on both
temperature $T$ and frequency $\omega$ associated with the presence
of FC, see Section \ref{Lin}. As a result, the scattering rate
becomes $1/\tau\simeq a_0+a_1T$. This result is in good agreement
with the experimental facts \cite{tomph,chen} presented in Fig.
\ref{fig9}. Thus, the FC theory successfully explains the behavior
of both the scattering rate and the resistivity, see Eqs.
\eqref{DR2} and \eqref{DR3} \cite{lett2010,kz,prb:2013}.

It is worth noting that in the case of HF metals and high-$T_c$
superconductors the scattering rate is a linear function of
$\omega$. Thus, we have to take into account the general expression
for the optical conductivity, since omitting the real part of the
scattering rate leads to the Kramers---Kronig violation
\cite{got,lit}. To restore the Kramers---Kronig relation,  we
employ the complex presentation of the scattering rate
\begin{equation}
\frac{1}{\tau(\omega)}\propto \left\{\eta\,\omega\ln
{\left|\omega\over \omega_c\right|}+i(a_0+a_2\omega)\right\},
\label{dampp}
\end{equation}
with $\eta$ denoting the volume in momentum space occupied by the
flat band, and $\varepsilon_c$ being a characteristic constant
\cite{kz,prb:2013}. Upon inserting Eq. \eqref{dampp} into Eq.
\eqref{DRUD}, we get
\begin{equation}
\sigma_{opt}=\frac{ne^2}{m}\frac{1}
{a_0+a_1T+a_2\omega-i\omega(1+\eta\ln
{\left|\omega\over\omega_c\right|})}. \label{dlog}
\end{equation}
Taking into account that the logarithm $\eta\ln(\omega/\omega_c)$
on the right hand side of Eq. \eqref{dlog} is a "slow" function of
its variable, we approximate $1+\eta\ln(\omega/\omega_c)$ by a
constant $c$, as it is done in the next Subsection 6.1. Our
calculations show that constant $c$ is a good approximation for the
logarithm.

\subsection{Scaling behavior of the real part
$\sigma^R_{opt}$ of the optical conductivity}\label{expf}

Now we are in position to consider the scaling behavior of
$\sigma_{opt}(\omega,T)$. In the present context, the HF compounds
are taken to represent strongly correlated Fermi systems as
realized in HF metals and high-$T_c$ superconductors. One can
expect that HF compounds with their extremely diverse composition
and microscopic structure would demonstrate very different
thermodynamic, transport, and relaxation properties. Upon inserting
Eq. \eqref{DR1} into Eq. \eqref{DRUD}, we obtain
\begin{equation}
\sigma_{opt}(\omega,T)=\frac{ne^2}{m}\frac{1}{a_0+a_1T+a_2\omega-i\omega}.
\label{DR4}
\end{equation}

\begin{figure}[!ht]
\begin{center}
\includegraphics [width=0.47\textwidth]{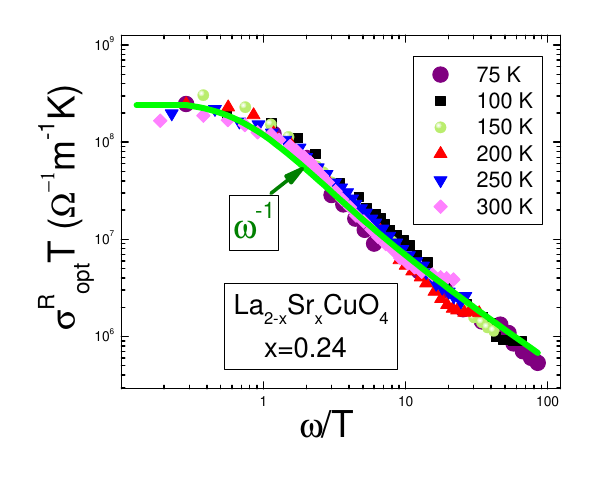}
\end{center}
\caption{The $\omega/T$ scaling behavior of $\sigma^R_{opt}T$ of
the cuprate $\rm La_{2-x}Sr_xCuO_4$ \cite{pasch}. At
$\omega/T\geq1$ $\sigma^R_{opt}\propto\omega^{-1}$, as it is shown
by the arrow. The solid curve is our theory. Here and below
theoretical real part $\sigma^R_{opt}(\omega,T)$ is given by Eq.\
(\ref{DR6}) with the parameters $b$ and $c$ chosen for the best
description of the whole set of experimental data. } \label{fig3}
\end{figure}

\begin{figure} [! ht]
\begin{center}
\includegraphics [width=0.47\textwidth]{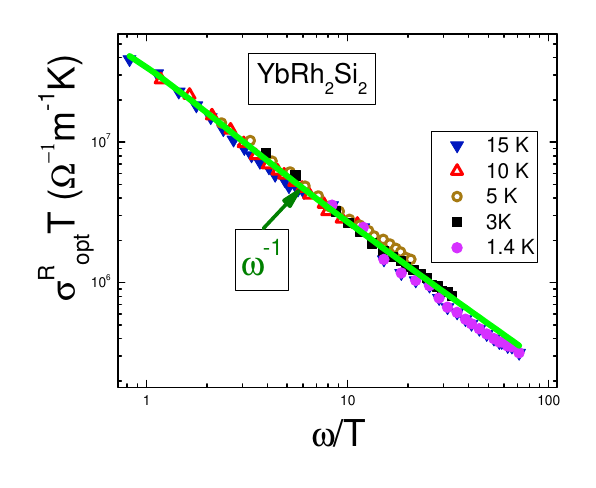}
\end{center}
\caption{The $\omega/T$ scaling behavior of the real part
$1/\sigma^R_{opt}T$ of the HF metal $\rm YbRh_2Si_2$ \cite{pasch}.
At $\omega/T>1$ $\sigma^R_{opt}\propto\omega^{-1}$, as it is shown
by the arrow. The solid curve is our theory.}\label{fig5}
\end{figure}

To compare our theoretical results with experimental facts, we
subtract the residual resistivity (or conductivity), as it is done
to the experimental facts \cite{homes,proch,mich,pasch}, and obtain
\begin{equation}
\sigma_{opt}(\omega,T)T=\frac{b(1+\omega/T)+ic\omega/T}
{b^2(1+\omega/T)^2+(c\omega/T)^2}, \label{DR5}
\end{equation}
where $b$ and $c$ are parameters. It is seen from Eq. \eqref{DR5}
that $\sigma_{opt}(\omega,T)T$ depends on the only variable
$\omega/T$. It directly follows from Eq. \eqref{DR5} that the real
part $\sigma^R_{opt}$ is given by
\begin{equation}
\sigma^R_{opt}T=\frac{b(1+\omega/T)}
{b^2(1+\omega/T)^2+(c\omega/T)^2}. \label{DR6}
\end{equation}

\begin{figure}[!ht]
\begin{center}
\includegraphics [width=0.47\textwidth]{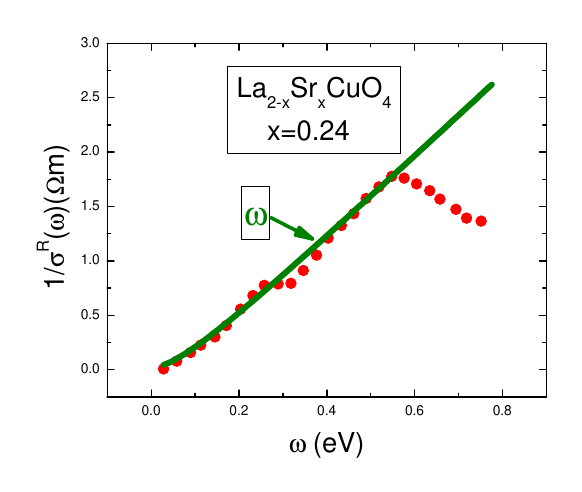}
\end{center}
\caption{Optical resistivity $1/\sigma^R_{opt}(\omega,T)$ at 75 K
of $\rm La_{2-x}Sr_xCuO_4$ with x=0.24, which is approximately
linear in frequency up to 0.6 eV \cite{pasch}. The solid line is
our theory.} \label{fig2}
\end{figure}

\begin{figure}
\begin{center}
\includegraphics [width=0.47\textwidth]{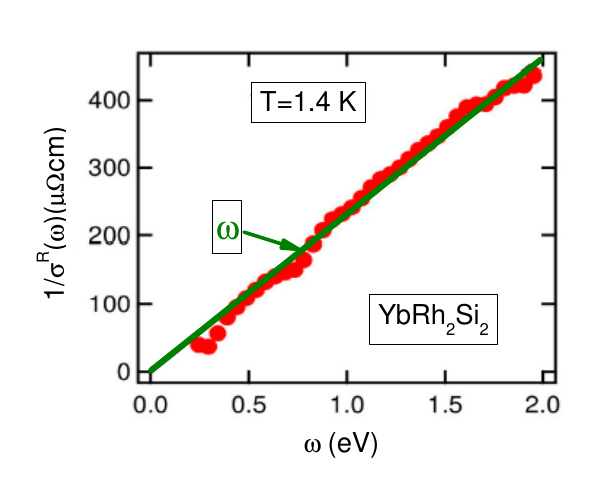}
\end{center}
\caption{Optical resistivity $1/\sigma^R_{opt}(\omega,T)$ at $1.4$
K of the HF metal $\rm YbRh_2Si_2$, which is approximately linear
in frequency \cite{pasch}. The solid line is our
theory.}\label{fig4}
\end{figure}

Figures \ref{fig3} and \ref{fig5} display the scaling of the
cuprate $\rm La_{2-x}Sr_xCuO_4$ and the HF metal $\rm YbRh_2Si_2$
in a wide range of the variable $\omega/T$. At $\omega/T>1$ the
real part is proportional $\omega^{-1}$,
$\sigma^R_{opt}\propto\omega^{-1}$, and demonstrates the NFL
behavior defined by the NFL behavior of $\tau$, see Eq.
\eqref{DR2}. Figures \ref{fig2} and \ref{fig4} show the optical
resistivity $1/\sigma^R_{opt}$ of $\rm La_{2-x}Sr_xCuO_4$ and the
HF metal $\rm YbRh_2Si_2$. It is seen that the resistivity is a
linear function of the variable, as it should be, and seen from
Figs. \ref{fig3} and \ref{fig5}.

The uniform scaling behavior seen from Figs. \ref{fig3} and
\ref{fig5} arises from the fact that HF compounds are located near
a topological FCQPT that generates their uniform scaling behavior
\cite{shagrep,book_20}. The emergence of the universal behavior,
exhibited by very distinctive HF metals, supports the conclusion
that HF metals represent a new state of matter
\cite{shag2017,book_20}. Unlike the situation of a conventional
quantum phase transition or the unconventional Kondo breakdown
scenario, the scaling induced by topological FCQPT, as seen in Fig.
\ref{fig3} and \ref{fig5}, occurs up to high temperatures, since
the behavior of the NPL is determined by quasiparticles, and not by
fluctuations or Kondo lattice effects \cite{shagrep,book_20}.

\subsection{Violation of the scaling behavior}

Now let us consider a possible violation of the observed scaling
behavior of optical conductivity, see Fig. \ref{fig3} and
\ref{fig5}. To understand the reasons for the violation, consider
the schematic phase diagram of HF compounds. At $T=0$ there is no
crossover region and the FC state is separated from the LFL region
by the first$-$order phase transition \cite{shagrep}, since the FC
state is characterized by a special quantum topological number,
being a new type of Fermi liquid \cite{volovik,volov_gr}. At $T>0$,
it is not a phase transition that occurs, but a crossover
\cite{shagrep}. At elevated magnetic fields reaching $B\geq T$, the
HF compound under consideration goes into the LFL state with
$\rho(T)\propto T^2$. As a result, we assume that both Eq.
\eqref{tu} and $\sigma^R_{opt}\propto\omega^{-2}$ become valid,
while the NFL behavior of $\sigma^R_{opt}\propto\omega^{-1}$
vanishes. Such a behavior can be observed in measurements of the
optical conductivity on the HF metal $\rm YbRh_2Si_2$ at low
temperatures under the application of magnetic field $B>B_{c0}$.
Here $B_{c0}\simeq$ 0.07 T is the magnetic field that tunes $\rm
YbRh_2Si_2$ to its antiferromagnetic quantum critical point
\cite{pfau,geg}.
\begin{figure}
\begin{center}
\includegraphics [width=0.47\textwidth]{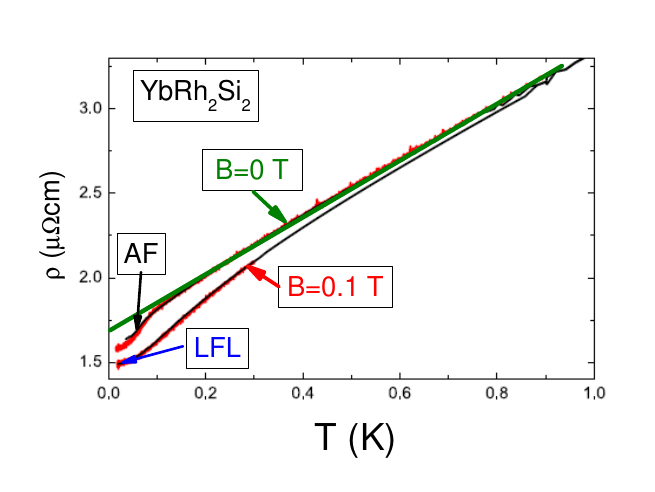}
\end{center}
\caption{Temperature dependency of the electrical resistivities of
$\rm YbRh_2Si_2$ single crystals at magnetic fields $B=0$ T and
$B=0.1$ T shown by the arrows \cite{pfau}. The antiferromagnetic
(AF) state at $B=0$ T and the LFL state at $B=0.1$ T are displayed
by the arrows.}\label{fig4a}
\end{figure}
It is seen from Fig. \ref{fig4a} that at $B>B_{co}$ and $\mu_BB\geq
k_BT$ $\rm YbRh_2Si_2$ exhibits the LFL behavior, as it does at low
temperatures in antiferromagnetic state \cite{geg}. The HF metal
$\rm YbRh_2Si_2$ is one of the purest HF metals. Therefore, the
regime of electron motion is ballistic. As a result, under the
application of weak magnetic field $B$ one could observe a positive
contribution $\delta\propto B^2$ to $\rho_0$ arising from orbital
motion of electrons induced by the Lorentz force. As seen from Fig.
\ref{fig4a}, $\rho_0$ diminishes, since the FC state is destroyed
by the application of magnetic field or by the antiferromagnetic
state, and $\rm YbRh_2Si_2$ demonstrates the LFL behavior with
diminishing the magnetoresistance, see Fig. \ref{fig4H}, while the
FC state itself creates the additional residual resistivity
$\rho_{FC}$ \cite{Khod_2020,kz}.
 {Thus, under the application of
magnetic field the LFL behavior is restored and the NFL one
exhibited by optical conductivity is violated, so that the scaling
behavior following from Eqs. \eqref{DR2} and \eqref{DR4} vanishes.
These measurement would confirm both our theoretical consideration
of the optical conductivity and the role of magnetic field when
studying the HF compounds \cite{mdpi23}. We note that this role of
magnetic field is missed in the frameworks of marginal Fermi
liquid, Kondo lattice, etc. \cite{shagrep,book_20}. }

\section{conclusion}\label{Concl}

In our short review, we have considered the transport properties of
HF metals and high-$T_c$ superconductors, and shown that transport
properties are defined by strong inter-particle interaction leading
to the topological FCQPT that forms flat bands, and makes the
linear in temperature resistivity, $\rho(T)\propto T$. We have
analyzed the magnetoresistance and shown that it under the
application of magnetic field becomes negative. We have shown that
the quasi-classical physics remains applicable to the description
of the resistivity $\rho\propto T$ of strongly correlated metals
due to the presence of a transverse zero-sound collective mode.
Thus, in the region of $T$-linear resistance, electron-phonon
scattering provides the lifetime $\tau$ of quasiparticles close to
material independence, which is expressed approximately through the
ratio of Planck's constant $\hbar$ to the Boltzmann constant $k_B$,
$T\tau\sim \hbar/k_B$. We have shown that due to the NFL behavior
of the resistivity $\rho(T)\propto T$, the real part
$\sigma^R_{opt}$ of the optical conductivity $\sigma_{opt}$
exhibits a similar NFL behavior $\sigma ^R_{opt}\propto\omega^{-1}$
rather than the well-known LFL relationship exhibited by ordinary
metals, $\sigma^R_{opt}\propto\omega^{-2}$. We have predicted that
under the application of magnetic field the real part of the
optical conductivity behave like
$\sigma^R_{opt}\propto\omega^{-2}$, since the corresponding HF
metal transits from the NFL behavior to the LFL one.

In summary, we have shown that the fermion condensation theory
provides a good description of the transport properties of various
HF compounds, as our results are in good agreement with
experimental observations.

\section{Acknowledgement} We thank V.A. Khodel for fruitful
discussions. This work was supported by U.S. DOE, Division of
Chemical Sciences, Office of Basic Energy Sciences, Office of
Energy Research, AFOSR.

\section{Author Contributions:}
V.R.S. and A.Z.M. designed the project and directed it with the
help of M.V.Z. V.R.S. and A.Z.M. wrote the manuscript and all
authors commented on it. The manuscript reflects the contributions
of all authors. All authors have read and agreed to the published
version of the manuscript.

\section{Institutional Review Board Statement:} Not applicable.

\section{Informed Consent Statement:} Not applicable. \section{Data Availability
Statement:} Not applicable.

\section{Funding:}
This research received no external funding.

\section{Declaration of interests:}
The authors declare no conflict of interest.

\end{document}